# Optical 3D cavity modes below the diffraction-limit using slow-wave surface-plasmon-polaritons

**Eyal Feigenbaum and Meir Orenstein**

*EE Department, Technion, Haifa 32000, Israel*
*meiro@ee.technion.ac.il*

**Abstract:** Modal volumes at the nano-scale, much smaller than the "diffraction-limit", with appreciable quality factors, are calculated for a dielectric cavity embedded in a space between metal plates. The modal field is bounded between the metal interfaces in one dimension and can be reduced in size almost indefinitely in this dimension. But more surprisingly, due to the "plasmonic" slow wave effect, this reduction is accompanied by a similar in-plane modal size reduction. Another interesting result is that higher order cavity modes exhibit lower radiation loss. The scheme is studied with effective index analysis, and validated by FDTD simulations.

## 1. Introduction

Cavities having nano-scale modal volume are very favorable for achieving significant nonlinear effects, low volume sensing, and strong material-light interactions. In the past few years, much attention was centered on achieving substantial light confinement with high quality factors by carefully designing microcavities [1] and defects in photonic crystals [2-4]. In 2D Photonic Band Gap (PBG) based cavities, high qualities were recently achieved by employing configurations that minimize the vertical scattering [5]. However, the minimal modal volume is basically bound in all these configurations to above $(\lambda/2n)^3$ – where $\lambda_0$ is the light wavelength in vacuum and n is the medium index of refraction .

    2D structures that include metal interfaces are known to relax the 1D confinement limitation. For instance, a mode in a dielectric-gap between two metal interfaces can be reduced in size almost indefinitely by reducing the gap width [6]. Although calculations show that the mode size may shrink to zero for vanishing interface spacing, we restrict the statement to "almost indefinite confinement" since the macroscopic Maxwell equations may fail at the few nanometers scale. The mode described above is confined only in one dimension and is unlimited in the other two. Exercising the same idea to achieve confinement in two or three dimensions (enclosing the dielectric cavity by metal) fails due to the emergence of a cut-off for the optical mode. Moreover, the incorporation of additional metal interfaces introduces more losses resulting in poor Q-factors.

    We propose here a novel configuration comprised of a 3D dielectric cavity formed in a gap between two metallic layers, which for the optical regime are essentially surface plasmon-polariton (SPP) supporting media [7]. This structure is beneficial both for reducing the vertical leakage as well as allowing unlimited shrinkage of the modal volume. Here the dielectric cavity is specifically implemented as a cylinder, while a similar configuration with photonic band gap cavities was presented by us in [8]. The results are validated with Finite-Difference Time Domain (FDTD) numerical solutions of Maxwell equations. Two related results should be noted – one is an experimental result of a Fabry-Perot cavity in a very thin dielectric between metal boundaries [9], exhibited nanoscale 2D confined cavity modes, with quality factor of ~10, however with the third dimension significantly larger than the diffraction limit. The second result – not related to plasmons, is a calculation of the modal confinement in a dielectric slot based cavity [10]. In this configuration – which does not incorporate metal and no slow wave effects exist, real 3D mode confinement below the diffraction limit cannot be achieved (in the sense that each dimension is smaller than half of the wavelength).

## 2. Plasmonic cavity analysis

We start by providing a short conceptual interpretation of the mechanism allowing to overcome the conventional diffraction limits for a 3 dimensional (3D) cavity modes, assisted by the plasmonic characteristics. This rational is similar to the discussion by Takahara et. al. [11] on plasmon assisted dielectric waveguiding. In any dielectric structure (having positive real dielectric constants), and specifically in a dielectric optical cavity cladded by dielectric materials, the *k*-vector components must be all real at least in one region of space (usually in the core). Thus the available *k* values are bounded (from above), according to $k_x^2+k_y^2+k_z^2=k_0^2\varepsilon$ ($k_0$ the vacuum wave-vector and $\varepsilon$ the dielectric constant) which yields (by a simple Fourier analysis) a minimal spatial bound of $(\lambda/2n)^3$ – known as the diffraction limit. For a metal cladded structure – the plasmon polariton solution is a slow wave such that in all space regions (including the cavity core) at least one k-vector component is imaginary ($k_x$ for metal layers located at x=constant). Thus there is no bound to *k* (from above), $k_y^2+k_z^2-|k_x^2|=k_0^2\varepsilon$, and consequently there is no limit to the mode size reduction. Moreover, the inplane *k*-vector (y,z) is enhanced by increasing the vertical component (x), translated to reduced inplane dimensions by reducing the vertical size - which is opposing to the trend in regular dielectric structures. This unique characteristics, emphasizes that in order to obtain cavities of sub-diffraction modal size in all 3D, the metal interfaces are required only in 1D.

To implement a "plasmonic" cavity according to the above principle, we propose a scheme where a cylindrical dielectric resonator, surrounded by air, is sandwiched in between two metal plates, as illustrated in the inset Fig. 1(a). The "plasmonic" effect is introduced into the vertical dimension as "metal air-gap" configuration [6]. The modal size can be reduced almost indefinitely in this configuration, while a simple introduction of metal interfaces into more than one dimension will result in modal cutoff [12]. An additional merit for the suggested dielectric cavity sandwiched between metal plates is the elimination of vertical (out of plane) radiation or leakage [13], similar to a defect in a 3D photonic band gap. This is strictly correct when ignoring the metal loss, but also when loss is taken into account - reasonable cavity qualities can be obtained. Since no output is possible through the metal layers – the output power leaks by a small in-plane radial radiation.

Equipped with this basic concept and with closed form solutions of a metal-gap structure [6,8] for the symmetric plasmon-polariton mode which does not exhibit a cutoff even for a vanishingly thin dielectric layer, we analyze the dielectric cylinder in between two thick metal layers, by introducing the effective index of the vertical mode into the in-plane structure [14]. First, the propagation constant ($\beta$) of the metal-gap structure mode is resolved from the dispersion relations, inside and outside the dielectric cylinder (similar to [6]). Then, the modal effective index ($n_{eff}(x,z) = \beta(x,z)/k_0$) is assigned, leading to a 2D in-plane equivalent structure of a dielectric circular cavity. The plasmon-polariton solution is essentially TM and the magnetic field of the cavity mode is:

$$H_\theta^{m,l}(\theta,r) = e^{jm\theta} \begin{cases} J_m(k_0 n_1 r) & r \leq a \\ A_{m,l} H_m^{(2)}(k_0 n_2 r) & r \geq a \end{cases} \quad (1)$$

where $n_1$ and $n_2$ are effective indices of the cylindrical cavity core and cladding; $a$ is the cylinder radius; and $J_m$ and $H^{(2)}_m$ are Bessel and second kind Hankel functions of order m. The valid cavity modes, characterized by azimuthal and radial numbers $\{m,l\}$, have continuous $H_\theta$ and $\partial_r H_\theta$ at $r=a$. The boundary conditions, for a given in-plane cylinder and for an (m) azimuthal order, yield a set of discrete radial solutions (indexed by $l$) – each having a distinct modal frequency.

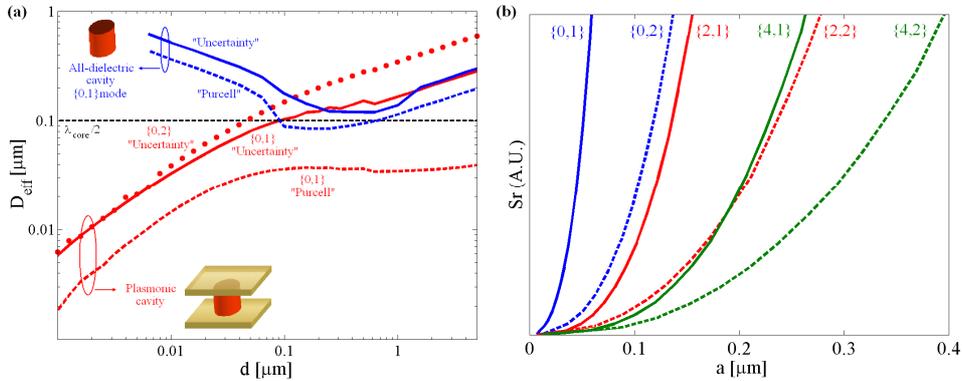

Fig. 1. Analysis results: (a) averaged modal size vs. cylinder height at $\lambda_0$=700nm, according to both our definition of "uncertainty volume" (Eq. 2) as well as the Purcell effective volume. (b) Outgoing radial power vs. cylinder radius for 20nm gap. $n_{si}$=3.5, $n_0$=1, $\lambda_{plasma}$=137nm.

The definition of modal volume is an involved issue. While most of the cavity related literature is using the effective mode volume obtained from the Purcell effect calculations [15], we believe that this effective mode volume (while being important for exhibiting the Purcell effect) is in fact the spatial intensity enhancement factor at the point of maximum intensity rather than the "actual mode volume". Furthermore – this measure has nothing to do with diffraction effects which are related to the "uncertainty" of the mode distribution, namely the intensity distribution variance. In addition – we do not expect that plasmonic nanocavities will serve for atomic physics (due to their relatively low Q), but rather be employed for small volume sensing, nano-lasers etc., where mode volume related to confinement factor is important. Since our main interest is the relation between plasmonic cavities and diffraction – we characterized our structure (Fig. 1a) by a mode "uncertainty" volume $V=D_{eff}^3=(\pi R_{eff}^2 h_{eff})$ where $R_{eff}$ and $h_{eff}$ are the variance of the intensity distribution along the radius and height respectively:

$$h_{eff} = 2\sigma_I = 2 \cdot \sqrt{VAR\{I\}} = 2\sqrt{\frac{\int x^2 |I(x)|^2 dx}{\int |I(x)|^2 dx}}; \quad R_{eff}^2 = \frac{\int r^2 |I(r)|^2 rdr}{\int |I(r)|^2 rdr} \qquad (2)$$

In order to facilitate the comparison with other reported cavities, we provide here also the Purcell effective mode volume (Fig. 1a) which always yields a significantly lower value.

Reducing the gap width (d), at a fixed wavelength, enhances the contrast between the effective indices of the dielectric core and clad, resulting in the reduction of the effective radius. This is just the opposite from the dielectric cylinder-slab case. The plasmonic configuration superiority is evident in Fig. 1(a), as the plasmonic cavity exhibits a substantial sub "diffraction limit" volume. More important is that there is no lower bound on the height for the metal embedded cylinder case (red). The averaged modal size of the regular dielectric slab (blue) is always above the "diffraction limit" (dashed-black). The calculations are illustrated for two modes, namely {m,l}={0,1} and {m,l}={0,2}, at a given wavelength of 700nm. The cylinder radius is varied to restore the cavity resonance for the given fixed wavelength. The averaged modal size is enhanced with modal order, in both configurations.

For a given structure excited with broadband pulse (as in the following Finite-Difference Time Domain - FDTD - simulations), the modal wavelengthes accommodate with the given cylinder radius. A unique feature in the plasmonic assisted cavity is that higher Q factors are expected for high order modes, which is counterintuitive with the common wisdom of regular dielectric cavities. For regular cavities, higher order in-plane modes have lower vertical *k*-vector, resulting in enhanced vertical radiation losses. For the "plasmonic" cavity, the metal layers disallow vertical radiation however they are a source for material loss. Here the inplane radiation losses of the mode into the dielectric clad, decrease with the order of the mode (Fig. 1(b)). Mathematically it can be traced to the fact that the solution outside the core (r>a), given in Eq. 1 (Hankel function of the second kind), vanishes rapidly for higher orders. On the other hand, the increased modal order is accompanied by a higher self-frequency, which suffers from larger material loss since it is closer to the plasma frequency. The interplay of these two mechanisms is expected to result in an optimized quality factor for modes of intermediate orders.

### 3. Simulative validation

To validate the predicted cavity performance, we performed FDTD based simulation, incorporating the complex metal dielectric function of the Drude model (which is a good approximation for our wavelengths). A silicon (or GaAs) cylinder 100nm in diameter was inserted into a 20nm gap between gold plates. A broadband pulse was introduced into the cavity from within (Gaussian shaped with x polarization) to allow for multiple resonances excitation. The resulting major cavity mode has a spectral peak of ~700nm, and we used this wavelength to excite the cavity by a CW excitation. The resulting vertical and in-plane distributions of the magnetic field are depicted in Fig. 2.

While the vertical distribution is confined to the 20nm inter-metal gap, the inplane field is confined inside the cylinder, exhibiting a major and two secondary lobes which are slowly rotating in time about the cylinder center during time evolution. The intensity FWHM resolved effective modal volume of $(36nm)^3$ is more than a order of magnitude smaller than the "diffraction limit", and has a Q-factor of 170, calculated by the ratio of total stored energy in a volume encompassing the cavity to the total outgoing power from the volume surfaces. The Q-factor can be enhanced further by employing lower loss metals (e.g. silver) or for increased inter-metal gap.

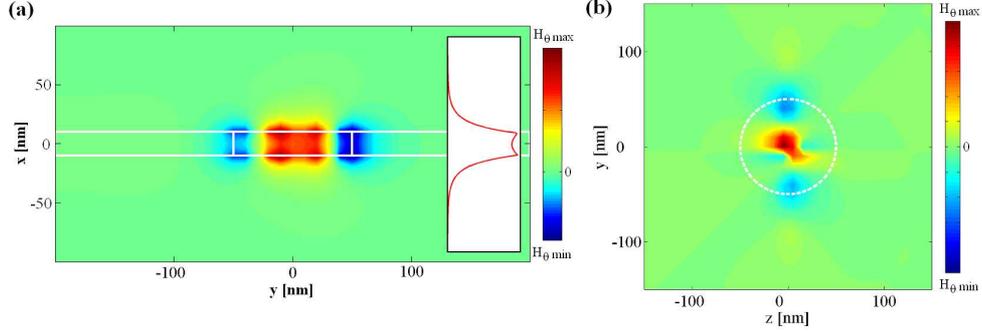

Fig. 2. FDTD simulations: Gold thickness: 100nm (practically infinite). Excitation: short x polarized pulse with Gaussian y distribution inside the cavity. In-plane and vertical resolutions are 10nm and 1nm respectively. Major spectral peak is at ~700nm. $H_\theta$ distribution: (a) vertical plane at z=0 (result obtained from CW excitation at 700nm for higher resolution). Inset: the field profile along y=20nm. (b) in-plane at x=0 (result obtained from the impulse excitation since CW has poorer visualization due to interference with the continuous source).

To better comprehend the in-plane field distribution we compared it to the results of the effective index analysis. In Fig. 3(b) the cylinder radius vs. the modal wavelength is depicted for an air-gap of 20nm between gold plates. For each cylinder radius and a given azimuthal order (m) the modal wavelength of the discrete radial solutions (*l*) is extracted. The balance between the vertical metal losses and in-plane radiation losses retains the {0,2} and {2,1} modes, both having similar eigen frequencies of about 700nm. Their interference matches the mode distribution and spectrum as obtained in the simulation and the expected axial rotation of the pattern is due to the slight difference in the modal propagation constants. The calculated field confinement (Fig. 3(a)) is slightly smaller compared to the simulation results - effective modal volume of $(41nm)^3$), which may be due to the influence of the imaginary part of the metal dielectric constant, not included in the effective index analysis.

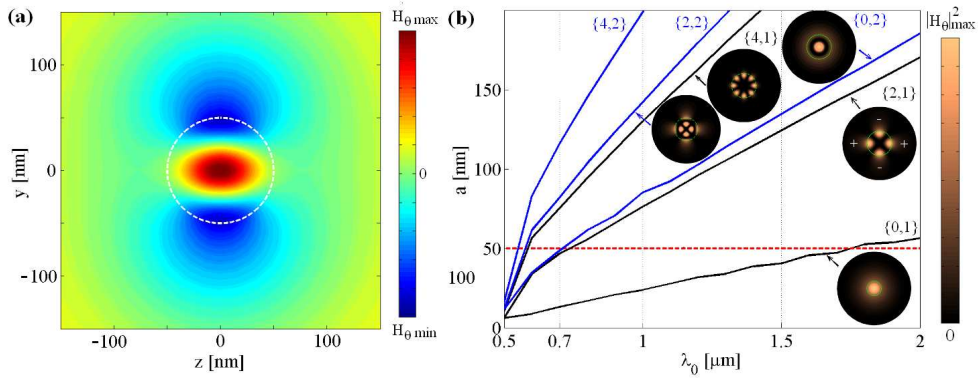

Fig. 3. Modal calculations by the effective index method for d=20nm: (a) coherent summation of modes {0,2}{2,1} at $\lambda_0$=700nm, a=50nm (cylinder boundaries in dashed white). (b) Cylinder radius supporting specific modes vs. the wavelength ($\lambda_0$). Red dashed line denotes the 50nm radius used in FDTD simulations. The inset shows the $|H_\theta|2$ distributions for the different modes for a=50nm (with respective resonance frequencies) (cylinder boundaries in dashed green).

## 4. Conclusion

We showed that incorporation of metal interfaces into a cavity scheme enables the emergence of cavity modes having nano-scale volumes, much smaller than the "diffraction limit", with moderate Q-factor. Introducing a dielectric cavity in between metal plates enables the suppression of vertical radiation and an almost indefinite vertical compression of the mode by reducing the gap width. The "plasmonic" slow wave effect results in a corresponding in-plane modal-size reduction. The effective index analysis is validated both quantitatively and qualitatively by FDTD simulations, which verifies also that due to higher outgoing radiation of lower order modes – intermediate order modes have higher Q factors.

**Acknowledgements**

We would like to acknowledge the Israel Ministry of Science and Technology for a partial support of this research.